\documentclass{jpp_acc}
\usepackage{graphicx}
\usepackage[utf8]{inputenc}
\usepackage[T1]{fontenc}
\usepackage{amsmath}
\usepackage{amsfonts}
\usepackage{amssymb}
\usepackage{newtxmath}
\usepackage[breaklinks=true,colorlinks=true,linkcolor=blue,urlcolor=blue,citecolor=blue]{hyperref}

\shorttitle{On the control of electron heating for optimal laser radiation pressure ion acceleration}
\shortauthor{H.-G. J. Chou, A. Grassi, S. H. Glenzer, and F. Fiuza}

\title{On the control of electron heating for optimal laser radiation pressure ion acceleration}

\author{
    H.-G. Jason Chou\aff{1,2}
    \corresp{\email{jasonhc@slac.stanford.edu}},Anna Grassi\aff{1},
    Siegfried H. Glenzer\aff{1},
    \and Frederico Fiuza\aff{1}
    \corresp{\email{fiuza@slac.stanford.edu}}
}

\affiliation{
\aff{1}High Energy Density Science Division, SLAC National Accelerator Laboratory, Menlo Park, CA 94025, USA
\aff{2}Department of Physics, Stanford University, Stanford, CA 94305, USA
}

\begin{document}

\maketitle

\begin{abstract}
We study the onset of electron heating in intense laser-solid interactions and its impact on the spectral quality of radiation pressure accelerated ions in both hole boring and light sail regimes. Two- and three-dimensional particle-in-cell (PIC) simulations are performed over a wide range of laser and target parameters and reveal how the pulse duration, profile, polarization, and target surface stability control the electron heating, the dominant ion acceleration mechanisms, and the ion spectra. We find that the onset of strong electron heating is associated with the growth of the Rayleigh-Taylor-like instability at the front surface and must be controlled to produce high-quality ion beams, even when circularly polarized lasers are employed. We define a threshold condition for the maximum duration of the laser pulse that allows mitigation of electron heating and radiation pressure acceleration of narrow energy spread ion beams. The model is validated by three-dimensional PIC simulations, and the few experimental studies that reported low energy spread radiation pressure accelerated ion beams appear to meet the derived criteria. The understanding provided by our work will be important in guiding future experimental developments, for example for the ultrashort laser pulses becoming available at state-of-the-art laser facilities, for which we predict that proton beams with $\sim$150 - 250 MeV, $\sim$30\% energy spread, and a total laser-to-proton conversion efficiency of $\sim$20\% can be produced.
\end{abstract}

\section{Introduction}\label{sec:intro}
In the last two decades, there has been a significant effort in exploring the generation of high-energy (1 -- 100 MeV) ion beams in plasmas produced by intense ($I>10^{18}$ W/cm$^{2}$) laser-solid interactions \citep{Daido2012,Macchi2013}. It has been shown that accelerating gradients as high as TV/m can be created in the plasma \citep{Wilks2001EnergeticInteractions} and ion beams can be produced with small emittance ($<0.1 \pi$ mm-mrad; \citealt{Borghesi2004Multi-MeVInteractions,Cowan2004UltralowAccelerator}) and short bunch duration ($\lesssim1$ ps; \citealt{Dromey2016PicosecondBursts}). These promising results and the potential to produce high-energy, high-current ion beams in more compact systems than solid-state based accelerators \citep{Cahill2018HighCavities} make the study of laser-driven ion acceleration an active area of research.

Laser-driven ion beams are now routinely used for radiography of high-energy-density plasmas \citep{borg2002,Rygg2008ProtonImplosions.} and hold promise for applications in isochoric heating of materials \citep{Patel2003IsochoricBeam,Tahir2005ProposalDarmstadt}, fast ignition of inertial confinement fusion \citep{roth2001}, injectors for conventional accelerators \citep{Antici2008NumericalSource,Aymar2020}, and tumor therapy \citep{Bulanov2008AcceleratingPulses,Kraft2010Dose-dependentBeams,Loeffler2013ChargedDirections,Bulanov2014LaserTherapy,Linz2016Laser-drivenRevisited,Kroll2022}. Important requirements for many of these applications are the ability to produce controllable, quasi mono-energetic (low energy spread) ion beams and at a high-repetition rate. These remain significant challenges for laser-driven ion beams, despite the progress in exploring different acceleration mechanisms, and significant developments in high-repetition rate targets \citep{Kim2016DevelopmentExperiments,Gauthier2017HighJets,Gode2017RelativisticInteractions,Obst2017EfficientJets,Curry2020CryogenicScience}.

The most studied laser-driven ion acceleration mechanism to date --- both theoretically and experimentally --- is the target normal sheath acceleration (TNSA; \citealt{Snavely2000IntenseSolids,Wilks2001EnergeticInteractions,Mora2003PlasmaVacuum}). Hot electrons are produced near the front surface of the target during the laser-plasma interaction, typically via $\mathbf{J}\times\mathbf{B}$ \citep{Kruer1985JxBLight,May2011} and Brunel (vacuum) heating \citep{Brunel1987Not-so-resonantAbsorption} mechanisms, which are maximized for linearly polarized lasers \citep{Wilks1997,Gibbon2005ShortMatter}. These hot electrons cross the dense target and escape into the vacuum on the rear side, setting up a strong space-charge sheath field that accelerates the target ions from the back surface in the target-normal direction. Ion beams produced by this mechanism are laminar and possess small emittance, however their energy spectrum is very broad, being typically characterized by an exponentially decreasing energy distribution \citep{Wilks2001EnergeticInteractions, Snavely2000IntenseSolids, Mora2003PlasmaVacuum}. 

Alternative ion acceleration schemes have been proposed in order to produce more narrow (quasi mono-energetic) ion spectra, including collisionless shock acceleration (CSA; \citealt{Denavit1992AbsorptionTargets,Silva2004ProtonInteractions,Fiuza2012Laser-drivenBeams,Haberberger2012Collisionlessbeams}) and radiation pressure acceleration (RPA; \citealt{Wilks1992AbsorptionPulses,Esirkepov2004HighlyRegime,Macchi2005LaserPlasmas,Robinson2008RadiationPulses}). CSA relies on the reflection of ions off a moving electrostatic shock front produced by the laser-plasma interaction near the front surface, which travels at roughly a constant speed inside the target. A small fraction ($\simeq1 - 10\%$) of the bulk ions is reflected by the shock producing a narrow energy spread ion beam. Hot electrons are important for driving the ion-acoustic waves that mediate shock formation inside the target, and therefore TNSA will also naturally accompany CSA, which can inadvertently broaden the ion spectrum. Specific shaping of the target density has been proposed as a way to control TNSA and produce high-quality beams from CSA \citep{Fiuza2012Laser-drivenBeams,Fiuza2013IonShocks}. Recently, tuning of the plasma density profile using a second laser was shown to produce narrow energy spread ion beams from CSA \citep{Pak2018CollisionlessLasers}. However, achieving precise control of the plasma density profile remains a challenge.

RPA relies on the slowly varying, cycle-averaged component of the ponderomotive force exerted by the intense laser pulse on the electrons at the front surface of the solid target. It is this radiation pressure that creates a charge separation between the electrons and the ions that accelerates the latter. It can potentially result in the generation of a quasi mono-energetic and laminar ion beam, provided that the accelerating structure is maintained stable and with uniform velocity, similarly to the case of CSA. In theory, RPA can produce ion beams with very high density because almost all ions in the laser focal region can be accelerated by the space-charge field. However, the experimental characterization of this acceleration scheme and observation of narrow energy spread ion beams have been challenging \citep{Henig2009Radiation-PressurePulses,Palmer2011MonoenergeticShock,Kar2012IonPressure,Steinke2013,bin2015,scullion2017,McIlvenny2021}. An important difficulty relies on the requirement of low electron heating for efficient momentum transfer from the laser to the ions, and to avoid other competing ion acceleration mechanisms, such as TNSA and CSA, to develop and dominate.  Indeed, recent experiments producing nearly 100 MeV proton beams likely involved the combination of different acceleration schemes and the observed energy spectra were broad \citep{Kim2016RadiationPulses,Wagner2016MaximumTargets,Higginson2018Near-100Scheme,Shen2021ScalingFoils}. Other significant challenges include the mitigation of corrugations at the laser-target interaction surface arising due to instabilities \citep{Palmer2012Rayleigh-TaylorLaser,Eliasson2015InstabilityLaser,Sgattoni2015Laser-drivenStructures,Wan2020EffectsAcceleration,Chou2022} and finite laser spot effects \citep{Klimo2008MonoenergeticPulses,Dollar2012FiniteTargets}and the control of the pre-plasma level that is naturally formed from the pre-heating and expansion of the target by a laser pre-pulse, which poses significant constraints on the laser contrast \citep{Varmazyar2018}.

In all these laser-driven ion acceleration schemes electron heating plays a major role in controlling the dominant acceleration mechanism and the quality of the accelerated ion beams. While it is well established that the laser polarization --- linear versus circular --- can be important in controlling electron heating via the $\mathbf{J}\times\mathbf{B}$ mechanism (\emph{e.g.} \citealt{May2011}), it is not clear how, more generally, the different laser properties affect the interplay between competing processes and instabilities at the front surface of the target in order to ensure a robust control of electron heating and ion acceleration.

Here, we use two- (2D) and three-dimensional (3D) particle-in-cell (PIC) simulations with the fully relativistic electromagnetic code OSIRIS \citep{Fonseca2002,Fonseca2008One-to-oneSimulations,Fonseca2013ExploitingAccelerators} to investigate in detail how the laser and plasma properties determine electron heating and how this will impact the quality of the accelerated ion beams. We focus in particular on RPA, exploring both hole boring (HB) and light sail (LS) regimes. We identify the dominant processes and establish a set of criteria relating the laser and target parameters that enable robust mitigation of electron heating and acceleration of high quality ion beams. This work presents a more detailed analysis of recently published results on the optimization of LS ion acceleration \citep{Chou2022} and expands on it by presenting new results on the HB regime. The new understanding and set of conditions provided can have an important impact in the guiding of future experiments and in ensuring a better characterization of different ion acceleration regimes, by isolating the dominant mechanisms. 

This paper is organized as follows. The physical regimes considered and the simulation setup used in this study are described in Sec.~\ref{sec:pars}. In Sec.~\ref{sec:ppol}, we show that suppression, or significant mitigation of electron heating is required to produce high-quality ion beams based on HB or LS. When this is not achieved, for thick targets a collisionless shock is formed which, in combination with TNSA, supersedes HB, and, for thin targets LS gives way to rapid decompression and transparency of the target. In Sec.~\ref{sec:sc} we discuss how the deformation of the target surface due to the growth of instabilities and finite spot size effects controls electron heating. We show that the Rayleigh-Taylor-like instability (RTI) is dominant in determining the onset of strong electron heating and controlling the quality of the accelerated ion beams. Based on this understanding, we define a threshold condition for the duration of the laser pulse that allows mitigation of electron heating and high-quality ion acceleration in both HB and LS regimes. The effect of the laser temporal profile on ion acceleration is studied in Sec.~\ref{sec:profile}, where it is shown that the Gaussian temporal profile leads to an increase of the ion energy spread. In Sec.~\ref{sec:prepulse} we discuss the importance of controlling laser pre-pulse to limit the electron heating due to the formation of pre-plasma. In Sec.~\ref{sec:discussion} we demonstrate that the new pulse duration conditions indeed minimize the electron heating and the energy spread of the accelerated ion beams over a wide range of laser and target parameters. Furthermore, we show that when the new threshold for pulse duration is combined with the condition for the optimal target thickness for LS, it limits the maximum laser intensity that can be used and the maximum peak energy of the accelerated ion beam. In Sec.~\ref{sec:disc} we verify these findings with 3D PIC simulations and demonstrate that based on the established criteria it is possible to generate high-quality ion beams from RPA (in both HB and LS regimes) using realistic laser and target conditions. Finally, in Sec.~\ref{sec:con} we present our conclusions and discuss the implications of the work for experimental studies.

\section{Radiation pressure acceleration regimes and simulation setup}\label{sec:pars}
We consider an intense laser interacting with a planar target with density $n_0>n_c$ (\emph{i.e.,} an overdense target), where $n_0$ and $n_c=m_e\omega_0^2/4\pi e^2$ are the initial plasma density and the critical density associated with laser propagation in the plasma, with $e$ being the elementary charge, $m_e$ the electron mass, and $\omega_0$ the laser frequency. The intense laser exerts a coherent ponderomotive force on the electrons at the surface of the solid target, creating a charge separation between the pushed electrons and the heavier ions, which in turn accelerates the ions. In practice, the laser radiation pressure acts as a piston pushing the plasma at the front surface inwards as the laser light is reflected from the surface. Using a 1D model based on energy- and momentum-flux conservation at the target surface, one finds that the front surface is pushed at the known HB velocity \citep{Wilks1992AbsorptionPulses}
\begin{align}
\label{eqn:vhb}
\frac{v_\text{HB}}{c}=\sqrt {\frac{P_L}{2m_in_ic^2}} = \sqrt {\frac{1+R}{4}\frac{Z}{A} \frac{m_e}{m_p}\frac{n_c}{n_0}}a_0\cos\theta_0,
\end{align}
where $P_L = (1+R)I\cos^2\theta_0/c$ is the radiation pressure exerted by the laser on the target surface in the normal direction, $Z$ and $A$ are the ion charge and mass numbers, $m_i$ and $m_p$ are the ion and proton masses, $c$ the speed of light, $a_0 \simeq 0.85\sqrt{I [\text{W/cm}^{2}](\lambda_0[\mu\text{m}])^2/10^{18}}$ the peak normalized vector potential, $\theta_0$ the incidence angle, $\lambda_0$ the wavelength of the laser, and $R\leqslant$ 1 the reflection coefficient. We note that $R$ may be a function of $\theta_0$, $\lambda_0$, $a_0$, and target density and composition.

We distinguish between two different RPA regimes: HB \citep{Wilks1992AbsorptionPulses,Macchi2005LaserPlasmas} and LS \citep{Esirkepov2004HighlyRegime,Robinson2008RadiationPulses,Macchi2010IonTargets}. The acceleration regime is determined by the ratio $l_0/(v_{\rm HB} \tau_0)$, where $l_0$ is the target thickness and $\tau_0$ is the laser pulse duration. In the HB regime, $l_0 > v_\text{HB} \tau_0$ and the laser radiation pressure can only push a small fraction of the target. Ions are reflected once off the laser piston acquiring a velocity $v_i \simeq 2v_\text{HB}$, or equivalently a peak energy per nucleon $\epsilon_0 = 2m_pv_\text{HB}^2$. In the LS regime, $l_0 < v_\text{HB} \tau_0$ and the laser can push the whole target repeatedly during the laser pulse duration. In other words, the target is accelerated via multiple HB stages \citep{Macchi2010IonTargets,Grech2011EnergyBeams}. In this case, the target experiences an acceleration \citep{Macchi2010IonTargets}
\begin{align}
\label{eqn:als}
a_{\rm RPA} = \frac{d}{dt}(v_i\gamma_i) \simeq \frac{2I}{m_in_il_0c}R\frac{1-\beta_i}{1+\beta_i} \simeq 2v_\text{HB}^2/l_0,
\end{align}
where $\beta_i=v_i/c$ and $\gamma_i$ are the normalized ion velocity and Lorentz factor. The last equality is the approximation in the non-relativistic limit. For negligible laser electron heating ($R\simeq 1$), an exact solution for the final ion velocity $\beta_{i,0}$ and the corresponding energy per nucleon $\epsilon_0$ exists \citep{Macchi2009LightReexamined}: 

    \begin{align}
    \label{eqn:beta}
    \beta_{i,0}=\frac{(1+\xi)^2-1}{(1+\xi)^2+1};\     \epsilon_0=m_pc^2\frac{\xi^2}{2(\xi+1)},
    \end{align}
where  $\xi=c\frac{Zm_en_c}{Am_pn_0}\frac{a_0^2\tau_0}{l_0}$.

In our simulations a laser pulse with frequency $\omega_0$ is launched along the $x_1$ direction (which is also the direction of the target normal) from the left boundary and irradiates, unless otherwise stated, an electron-proton plasma (\emph{i.e.} $m_i=m_p=$ 1836 $m_e$) with initial density $n_0\geqslant$ 40 $n_c$. A minimum density of 40 $n_c$ corresponds to that of high-repetition rate liquid hydrogen targets  \citep{Kim2016DevelopmentExperiments,Gauthier2017HighJets,Curry2020CryogenicScience}. The plasma density follows a step-like profile with thickness $l_0$. An initial electron temperature $T_e=100$ eV is used (we have checked that our results are not sensitive to the initial temperature choice in the $10 - 1000\,$eV range). The typical size of the simulation box in 2D (3D) is 400 (300)$\,c/\omega_0$ longitudinally, and $250\,c/\omega_0$ transversely in $x_2$ (and in $x_3$). The 2D (3D) simulations use 16 (8) particles per cell per species and a spatial resolution of 0.2 (0.5)$\,c/\omega_{pe}$ in each direction, where $\omega_{pe}=\sqrt{4\pi e^2n_0/m_e}$ is the electron plasma frequency. The time step is chosen according to the Courant–Friedrichs–Lewy condition. Open (absorbing) boundary conditions for both particles and fields are used in the longitudinal and transverse directions (except in the cases with a plane wave laser where the transverse boundary conditions are periodic). We have tested different resolutions and numbers of particles per cell to ensure convergence of the results and have used a third order particle interpolation scheme for improved numerical accuracy. We have also tested different domain sizes to ensure that this domain allows capturing the electron heating and ion acceleration dynamics without the build up of fields at the boundaries due to the absorption of current from escaping particles. Because in most practical applications the primary interest is in highly directional ion beams, for simulations with a finite laser spot the ion energy spectra are integrated within a $10^\circ$ opening angle from the laser propagation (forward) direction. We have checked that this is consistent with selecting the ions within the area of the focal spot.

To study in detail how the laser-plasma parameters affect electron heating and ion acceleration, we have performed a parameter scan in laser intensity ($a_0=5 - 200$), polarization (P-, S-, and circular), incidence angle ($\theta_0 = 0 - 45^\circ$), Full-Width-Half-Maximum (FWHM) duration ($\tau_0 = 30 - 2000\,\omega_0^{-1}$), 
focal spot (at $1/e^2$ beam width; $w_0=4 - 50\,c/\omega_0$ and plane wave), target composition $1\leqslant A/Z\leqslant 4$ for single-species ions, and CH, density ($n_0 = 40 - 500\,n_c$, covering the range from liquid hydrogen to solid-density targets), and thickness ($l_0 = 0.08 - 40\,c/\omega_0$).

\section{Termination of radiation pressure acceleration due to strong electron heating}\label{sec:ppol}

    \begin{figure}
    \centering
    \includegraphics[width=.99\textwidth]{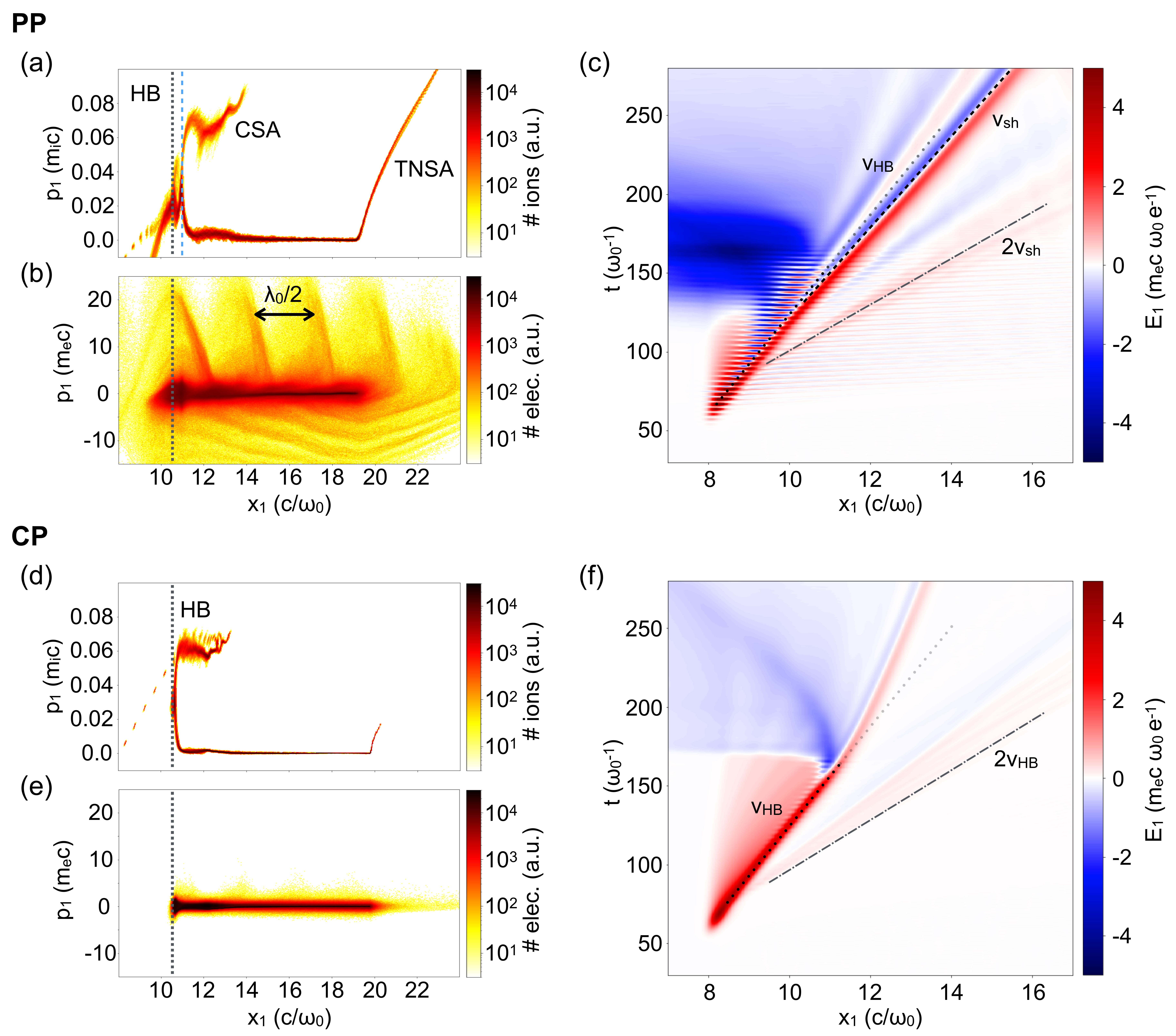}
    \caption{Results of 2D PIC simulations of a P-polarized (PP) [(a)--(c)] and circularly polarized (CP) [(d)--(f)] short-pulse laser ($a_0 = 12$) interacting with an overdense target ($n_0 = 42\,n_c$). Longitudinal $p_1-x_1$ phase spaces, shown at $t=150\,\omega_0^{-1}$, of protons [(a),(d)], electrons [(b),(e)], and time evolution of the longitudinal $E_1$ electric field [(c), (f)]. The laser pulse irradiates the target from the left-hand side, has a supergaussian temporal profile, and ends at $t\simeq160\,\omega_0^{-1}$. For a PP laser, the electrons are heated by the laser, HB is weakened, and a collisionless shock develops, which dominates the proton acceleration. The electrostatic shock front detaches from the HB front and propagates at $v_\text{sh}\simeq 0.035\,c$, which reflects the protons to a speed of $2v_\text{sh}$. The electrons in the CP case remain relatively cold and HB is the dominant ion acceleration mechanism, accelerating protons to $2v_\text{HB}$, where $v_\text{HB}\simeq0.031\,c$. After the laser ends, the HB front decays and slows down. In the phase spaces, the dotted lines denote the target front surface (with density $n \simeq n_c$) and the dashed blue line indicates the collisionless shock front. In (c) and (f), the dotted, dashed, and dash-dotted lines indicate $v_\text{HB}$, $v_\text{sh}$ and the proton beam velocity ($2 v_\text{HB}$ or $2 v_\text{sh}$) respectively.
    }
    \label{fig:ppol}
    \end{figure}
    
In this section, we show that for thick targets (HB regime) when there is significant electron heating at the target surface, HB gives rise to the formation of a collisionless shock that is launched into the target and dominates ion acceleration. We illustrate these results with 2D simulations where an intense ($a_0 = 12$) laser interacts with an overdense thick target ($n_0 = 42\,n_c$ and $l_0 = 12\,c/\omega_0$). The laser is either a P-polarized (PP, with the electric field along the $x_2$ direction) or circularly polarized (CP) plane wave, with a 4th-order supergaussian temporal profile with $\tau_0 = 100\,\omega_0^{-1}$.

In Figure~\ref{fig:ppol}(a) and (b), we show the longitudinal phase spaces of electrons and protons for the PP simulation near the end of the laser-plasma interaction at $t = 150\,\omega_0^{-1}$. We can clearly see hot electron bunches produced at a frequency of $2\omega_0$, which is a signature of the $\mathbf{J}\times\mathbf{B}$ heating mechanism \citep{May2011}. As a result, a significant fraction of the laser energy goes into the electron population, weakening HB. Indeed, the measured HB velocity is $v_\text{HB}\simeq 0.026\,c$, consistent with a laser absorption into hot electrons of $\simeq 50\%$ ($R \simeq 0.5$ in Eq.~\ref{eqn:vhb}). In addition, we observe that a collisionless shock forms and dominates proton acceleration as illustrated in Fig.~\ref{fig:ppol}(a) and (c). The shock front detaches from the surface and propagates into the target at a nearly constant velocity $v_\mathrm{sh} \simeq0.035\,c$. The proton population at $11\,c/\omega_0 \lesssim x_1\lesssim 14\,c/\omega_0$ has been accelerated by the shock front to $v_i\simeq2v_\text{sh}$. It is important to note that even after the laser interaction finishes, the collisionless shock continues to propagate through the target and reflect protons (Fig.~\ref{fig:ppol}c). It is also worth noting that due to the generation of hot electrons, a strong space-charge field develops at the rear surface of the target and leads to TNSA, which is evidenced by the proton phase space at $x_1\gtrsim 19\,c/\omega_0$. At later times, the protons accelerated by the shock acquire a large energy spread when they leave the target rear surface and experience TNSA. Tailoring of the rear side density profile is required to control TNSA and enable quasi mono-energetic ion beams from CSA \citep{Fiuza2012Laser-drivenBeams,Fiuza2013IonShocks}. In general, we have found that in configurations in which electrons become relativistic, CSA and TNSA will dominate the ion acceleration mechanisms over HB. This highlights the need to prevent or significantly mitigate electron heating in order to enable HB to be the dominant mechanism and to produce high-quality ion beams.

A commonly employed strategy to mitigate electron heating by both $\mathbf{J}\times\mathbf{B}$ and Brunel mechanisms in laser-plasma interactions is the use of a CP laser at near normal incidence ($\theta_0 \simeq 0^\circ$; \citealt{Macchi2005LaserPlasmas}). The $\mathbf{J}\times\mathbf{B}$ mechanism relies on the standing wave created by the incoming and reflected laser field. For linear polarization, the oscillation of the magnetic field at the surface allows for electrons to escape the target and experience the electric field of the laser, being accelerated transversely and then rotated back into the target by the magnetic field. For circular polarization, the magnetic field of the standing wave at the target surface does not decrease to zero --- it just rotates --- and thus electrons cannot escape the target to be efficiently accelerated \citep{May2011}. The Brunel heating mechanism relies on a laser electric field component normal to the surface to directly accelerate the electrons. This is absent for normal incidence, provided that the target surface remains uniform and stable. As we will discuss in more detail in the next section, these conditions can only be maintained for very short interaction times. Fig.~\ref{fig:ppol}(d)--(f) shows the results of a simulation with the same laser and plasma parameters as in Fig.~\ref{fig:ppol}(a)--(c) but using CP. We observe that CP is indeed capable of maintaining reduced electron heating, allowing HB to be the dominant ion acceleration mechanism. We observe that the HB velocity is $v_\text{HB} \simeq0.031\,c$, in good agreement with Eq.~\eqref{eqn:vhb} for $R \simeq 1$ and that the accelerated protons have $v_1\simeq 2v_\text{HB}$. We also find that, in contrast to the case of a collisionless shock, the HB velocity abruptly slows down when the laser-plasma interaction ends and proton reflection/acceleration ceases (Fig.~\ref{fig:ppol}f at t$\simeq 165\,\omega_0^{-1}$). This is an interesting difference between HB and CSA that impacts the total charge accelerated by each mechanism: while HB tends to reflect a larger fraction of the background ions, it can only do so during a shorter period when compared to CSA.

In addition to the use of a CP laser at near normal incidence ($\theta_0 \simeq 0^\circ$), it has been recently proposed in the context of collisionless shock studies \citep{Grassi2017Radiation-pressure-drivenShocks} that S-polarization (SP; with the electric field along the $x_3$ direction) with a large incidence angle can also achieve similar results in terms of mitigating electron heating. This is because Brunel heating is absent for SP and $\mathbf{J}\times\mathbf{B}$ heating can be significantly reduced since it scales with $\cos\theta_0$. We have explored this possibility over a large range of laser intensities ($a_0 = 5 - 30$) and plasma densities ($n_0 = 40 - 150\,n_c$) and confirmed that in 2D indeed SP with $\theta_0=45^\circ$ can significantly suppress electron heating and lead to HB ion acceleration comparable to the CP case (not shown here). However, the situation changes significantly for more realistic 3D simulations. In 3D, the laser-plasma interaction along the direction of laser polarization is effectively PP at $\theta_0 \simeq 0^\circ$ and gives rise to significant electron heating along the laser polarization. In all 3D SP cases tested, hot electrons acquire relativistic temperature and we observe a transition from HB to CSA similar to the PP case illustrated in Fig.~\ref{fig:ppol}(a--f). 

For very thin targets, corresponding to the LS regime, we have observed similar results in terms of the mitigation of electron heating, which was achieved with a CP laser with $\theta_0 \simeq 0^\circ$ and led to the acceleration of ions with a narrow energy spread. We note that for PP and SP, the onset of strong electron heating does not lead to the formation of a collisionless shock. Instead, we observe that the thin target quickly becomes transparent to the laser and the radiation pressure is no longer efficient in accelerating ions. Overall, for both HB and LS regimes, we find that the use of CP is required to significantly mitigate electron heating and optimize RPA. Non-normal incidence angles can be useful for SP, but unfortunately seem to be only effective in 2D. In the remainder of this paper, we discuss in detail the impact of the laser and plasma parameters on electron heating with CP for both HB and LS regimes in order to understand the set of conditions required for high-quality ion acceleration.

\section{Development of surface corrugations}\label{sec:sc}
The interaction of an intense laser with an overdense target can significantly modify the shape of the target surface either due to the development of surface instabilities or from finite laser spot size effects. These modulate the surface density profile, which can ultimately trigger strong electron heating (\emph{e.g.} \citealt{Klimo2008MonoenergeticPulses,Dollar2012FiniteTargets,Paradkar2016ElectronLaser}), even for a CP laser, and impact the quality and mechanisms of ion acceleration as discussed above. In this section, we discuss the importance of surface corrugations on HB and LS acceleration regimes with a CP laser. 

\subsection{Growth of surface instabilities}
\begin{figure}
    \centering
    \includegraphics[width=.95\textwidth]{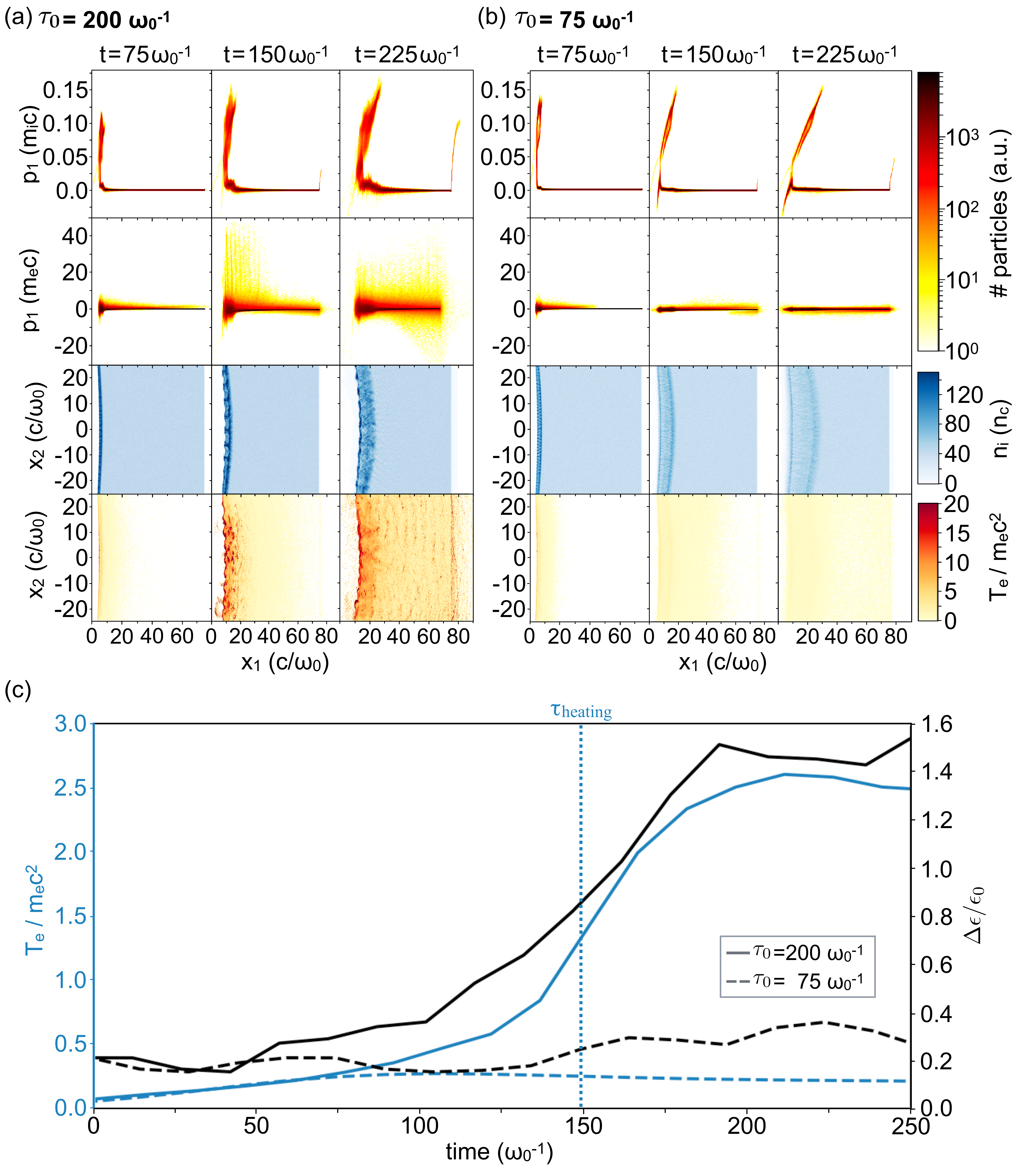}
    \caption{ Results of 2D PIC simulations of the interaction of an intense Gaussian CP laser pulse with an overdense target in HB regime. [(a),(b)] Longitudinal $p_1-x_1$ ion (top) and electron (second row) phase spaces, ion density profile (third row), and local electron temperature (bottom). The laser pulse durations are $\tau_0 = 200$ and $75\,\omega_0^{-1}$ in (a) and (b) respectively. (c) Temporal evolution of $T_e$ (blue, left axis) and ion beam energy spread $\Delta\epsilon/\epsilon_0$ (black, right axis). The time $t = 0$ is defined as $\tau_0/2$ before the laser peak intensity reaches the target. }
    \label{fig2}
    \end{figure}

\begin{figure}
    \centering
    \includegraphics[width=.95\textwidth]{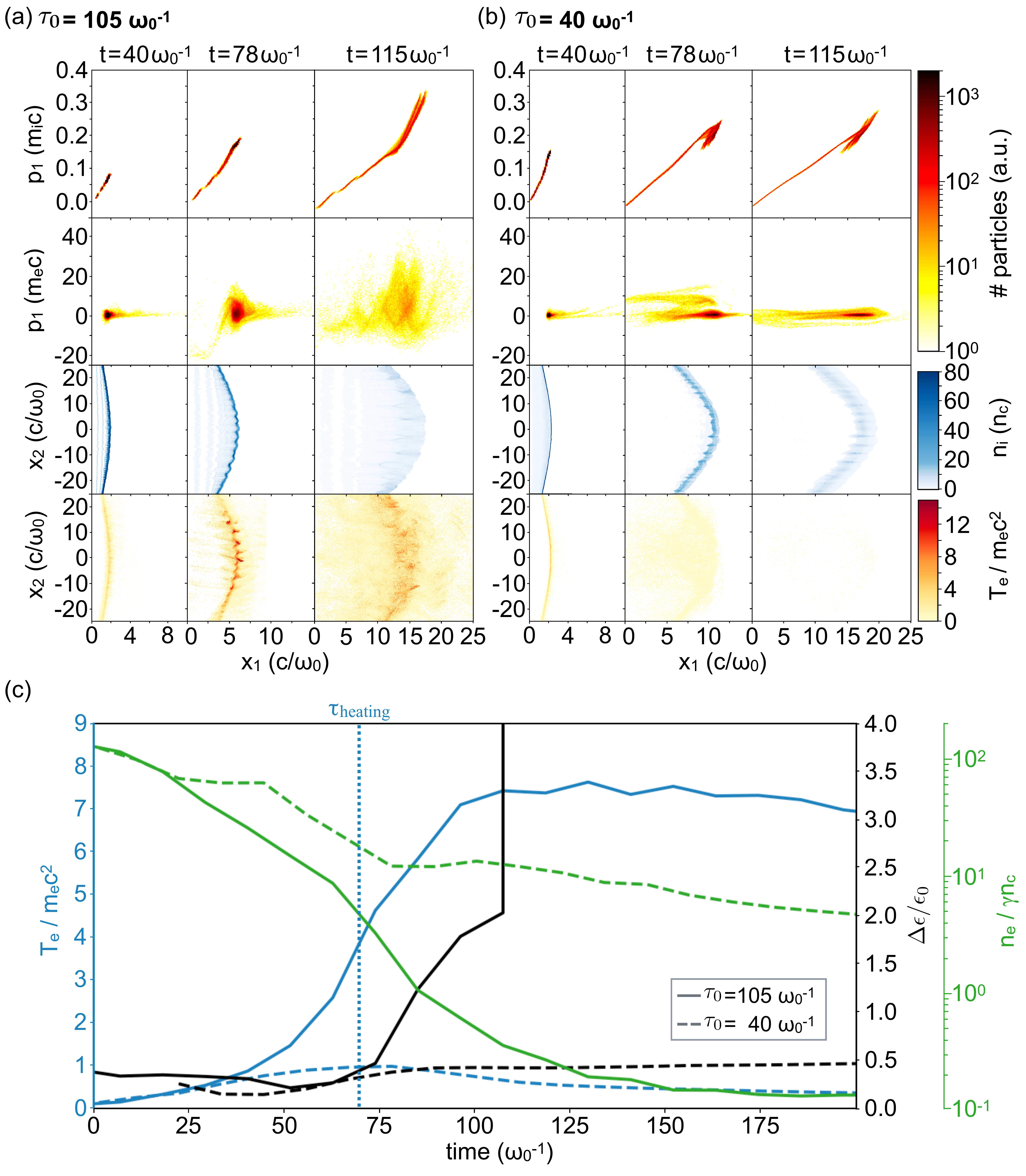}
    \caption{ Same as Fig.~\ref{fig2} but for LS, where the laser pulse durations are $\tau_0 = 105$ and $40\,\omega_0^{-1}$ in (a) and (b) respectively. In (c) the green, rightmost axis plots plots $n_e/(\gamma n_c)$. }
    \label{fig3}
    \end{figure}

Previous studies have investigated the development of density ripples at the interaction surface and there has been significant discussion on which instabilities are dominant, including the Weibel instability \citep{Sentoku2000MagneticPlasmas}, RTI  \citep{Gamaly1993InstabilityBeam,Pegoraro2007PhotonPulse,Palmer2012Rayleigh-TaylorLaser,Khudik2014TheStability,Eliasson2015InstabilityLaser,Sgattoni2015Laser-drivenStructures}, electron-ion coupling instabilities \citep{Wan2016PhysicalAcceleration,Wan2018PhysicalRegimes}, or a combination of these \citep{Wan2020EffectsAcceleration}. However, the correlation between these instabilities and the onset of electron heating has not been studied systematically for both HB and LS. Here, we begin by illustrating the surface dynamics and electron heating for the interaction of a CP laser with Gaussian longitudinal and transverse intensity profiles with a target at normal incidence using 2D PIC simulations. In the HB regime, we use a laser with $a_0=27$, $\tau_0=200\,\omega_0^{-1}$, $w_0=50\,c/\omega_0$ with $n_0 = 40\,n_c$ and $l_0=75\,c/\omega_0$; for LS, $a_0=15$, $\tau_0=105\,\omega_0^{-1}$, $w_0=50\,c/\omega_0$, with $n_0 = 250\,n_c$, and $l_0 =0.085\,c/\omega_0$ are used. The results are shown in Fig.~\ref{fig2}(a) for HB and Fig.~\ref{fig3}(a) for LS respectively.

We observe that indeed, even with CP, there is the onset of strong electron heating after a relatively short interaction time. In the HB regime (Fig.~\ref{fig2}a), the electron and ion phase spaces at $t=150\,\omega_0^{-1}$ show that electrons have been heated to relativistic temperatures, which leads to a significant weakening of the laser radiation pressure and to the formation of a collisionless shock and development of strong TNSA field that dominate ion acceleration, similarly to what was observed for the PP case discussed in Sec.~\ref{sec:ppol}. A shapr increase of the FWHM energy spread of the accelerated ions ensues, reaching $\Delta\epsilon/\epsilon_0 > 100\%$ as can be seen in Fig.~\ref{fig2}(c); and by $t = 225\,\omega_0^{-1}$ both the electron temperature and ion energy spread have saturated at large values.

In the LS regime (Fig.~\ref{fig3}a), we observe that once the electrons are significantly heated they drive the rapid expansion of the target and broadening of the ion energy spread (Fig.~\ref{fig3}c) due to the associated strong space-charge field and short target thickness. By $t = 115\,\omega_0^{-1}$ the target expansion leads to the onset of relativistic transparency as can be seen in Fig. \ref{fig3}(c) from the evolution of $n_e/(\gamma n_c)$ (the ratio of the electron density, $n_e$, and relativistic critical density at the target front surface, where $\gamma$ is the average Lorentz factor of the electrons). At this point, RPA is terminated and the ion energy distribution ceases to be peaked. 

From the temporal evolution of the electron temperature $T_e$ (average kinetic energy of electrons) and FWHM ion energy spread $\Delta\epsilon/\epsilon_0$ shown in Figs.~\ref{fig2}(c) and~\ref{fig3}(c) for the HB and LS regimes, respectively, we observe that they follow a similar behavior with the ion beam energy spread increasing sharply following the rapid growth of $T_e$ in both cases. We define this time associated with the onset of strong electron heating, $\tau_\text{heating}$, as the time for which the rate of increase of the electron temperature, $dT_e/dt$, is maximum.

    \begin{figure}
    \centering
    \includegraphics[width=.9\textwidth]{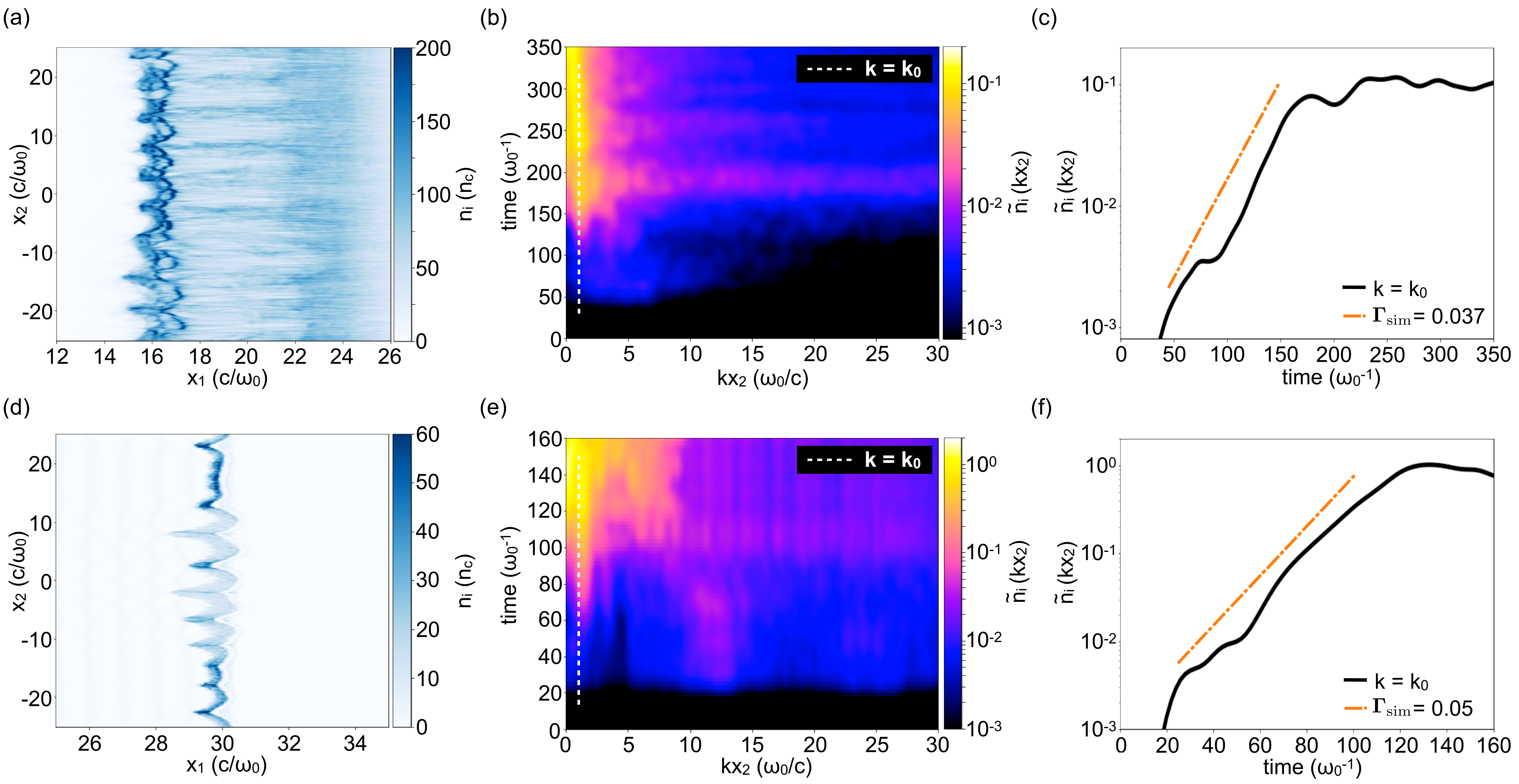}
    \caption{Development of surface corrugations in HB [(a)--(c)] and LS [(d)--(f)] regimes. [(a),(d)] Proton density showing transverse density corrugations near the interaction surface at $t = 200\,\omega_0^{-1}$ for (a) and $t = 90\,\omega_0^{-1}$ for (d). [(b),(e)] Evolution of Fourier modes at the relativistic critical surface of the corrugation amplitudes $\widetilde{n}_i(k_{x_2})$. The dashed white line denotes the $k_{x_2}=k_0$ mode. [(c),(f)] Time evolution of the $k_{x_2}=k_0$ mode and its linear fit.}
    \label{fig:sc}
    \end{figure}

We have repeated the same simulations in both regimes but using a shorter laser pulse with $\tau_0 = 75\,\omega_0^{-1}$ and $\tau_0 =  40\,\omega_0^{-1}$, both $< \tau_\text{heating}$, for HB and LS, respectively, to confirm the impact of electron heating on the growth of the ion energy spread and overall target dynamics. The results are shown in Figs.~\ref{fig2}(b,c) and~\ref{fig3}(b,c). In these cases, we observe that indeed both $T_e$ and $\Delta\epsilon/\epsilon_0$ remain low (Figs.~\ref{fig2}c and~\ref{fig3}c). The ion beam energy spread saturates at $t \simeq 2 \tau_0$ with $\Delta\epsilon/\epsilon_0 \ll 1$ and remains stable long after the laser-plasma interaction has finished.

We have found that the onset of electron heating in both HB and LS regimes is related to the emergence of large transverse density modulations at the target front surface with a wavelength comparable to that of the laser. For the longer pulse simulations ($\tau_0 > \tau_\text{heating}$), these surface density modulations are visible in the third rows of Figs.~\ref{fig2}(a) and~\ref{fig3}(a) at the times where electron heating is also observed. Density modulations with a wavelength comparable to $\lambda_0$ allow the penetration of the laser in the lower density regions and resonant enhancement of its electric field \citep{Eliasson2015InstabilityLaser,Sgattoni2015Laser-drivenStructures}, giving rise to effective electron heating, for example via the Brunel mechanism (bottom rows of Figs.~\ref{fig2}a and~\ref{fig3}a), with the temperature reached being comparable to that observed in simulations with a linearly polarized laser (not shown here). The spatial distribution of $T_e$ shows indeed that the heating is happening at the walls of these concave valleys and consistent with direct acceleration by the laser electric field. The location of the hot spots of $T_e$ oscillates (from top to bottom of the valleys) in accordance with the phase of the laser electric field. In the simulations with $\tau_0 < \tau_\text{heating}$, the amplitudes of the density modulations at the surface are much smaller during the time of laser interaction (third rows of Figs.~\ref{fig2}b and~\ref{fig3}b) leading to a much reduced electron heating and stable ion acceleration.

In order to study the mechanism responsible for these corrugations, and isolate the effects of surface instabilities, we have performed a parameter scan of 2D simulations with a long, plane-wave CP laser at normal incidence, with parameters varied in the following ranges: for HB, $5 \leqslant a_0 \leqslant 60$, 40 $n_c \leqslant n_0 \leqslant 200\,n_c$; for LS, $5 \leqslant a_0 \leqslant 200$, 40 $n_c \leqslant n_0 \leqslant 500\,n_c$, and $0.08\,c/\omega_0 \leqslant l_0 \leqslant 2\,c/\omega_0$. Note that for the LS regime, the initial target thickness $l_0$ is always larger than or equal to the optimal LS target thickness $l_\text{opt}$, defined as $l_\text{opt}= a_0\lambda_0n_c/(\sqrt{2}\pi n_0)$, for which the acceleration is maximized by minimizing the total target mass while guaranteeing that the target remains relativistically opaque \citep{Macchi2009LightReexamined}. For both regimes, the target is either composed of single-species ions with $1\leqslant A/Z \leqslant 4$, or CH (plastic). We analyze the growth of ion density modulations by computing the transverse Fourier modes ($k_{x_2}$) of the longitudinal ($x_1$) displacement of the relativistic critical surface ($n \simeq \gamma_0n_c$; where $\gamma_0=\sqrt{1+a_0^2/2}$ is the electron Lorentz factor), as a function of the transverse ($x_2$) position (\emph{e.g.} Fig.~\ref{fig:sc}b,e). Previous theoretical studies of surface instabilities considered perturbations of the surface displacement and showed it will grow exponentially due to instability (\emph{e.g.} \citealt{Gamaly1993InstabilityBeam,Eliasson2015InstabilityLaser}). Alternatively, for LS, one could obtain the Fourier modes of the amplitude of the transverse density profile by integrating over the target longitudinally. We have checked that the obtained growth rates and time scales of the different modes are consistent between both methods.

Figure~\ref{fig:sc} illustrates the growth of different modes for simulations with the same laser intensity and target parameters of Fig.~\ref{fig2} and~\ref{fig3}. Note that both targets remain opaque to the laser during the time of the analysis and thus the measurements of the growth rates and saturation levels are not affected by the onset of relativistic transparency. The fastest growing modes are observed at $k_{x_2}\gg k_0$, with $k_0=2\pi/\lambda_0$ (\emph{e.g.} $k_{x_2}\simeq13\,\omega_0/c$ in Fig.~\ref{fig:sc}e at $t\simeq30\,\omega_0^{-1}$), and have been previously described as associated with electron-ion coupling instabilities \citep{Wan2016PhysicalAcceleration,Wan2018PhysicalRegimes,Wan2020EffectsAcceleration}. However, these modes saturate at relatively low amplitude and do not lead to significant electron heating. The dominant density modulations are associated with the mode with $k_{x_2}\simeq k_0$ and we observe the onset of strong electron heating during the linear growth and saturation of this mode. This suggests that the onset of strong electron heating is related to laser-driven RTI for which the dominant mode is $k_{x_2}\simeq k_0$ \citep{Eliasson2015InstabilityLaser,Sgattoni2015Laser-drivenStructures}.

\begin{figure}
    \centering
    \includegraphics[width=.9\textwidth]{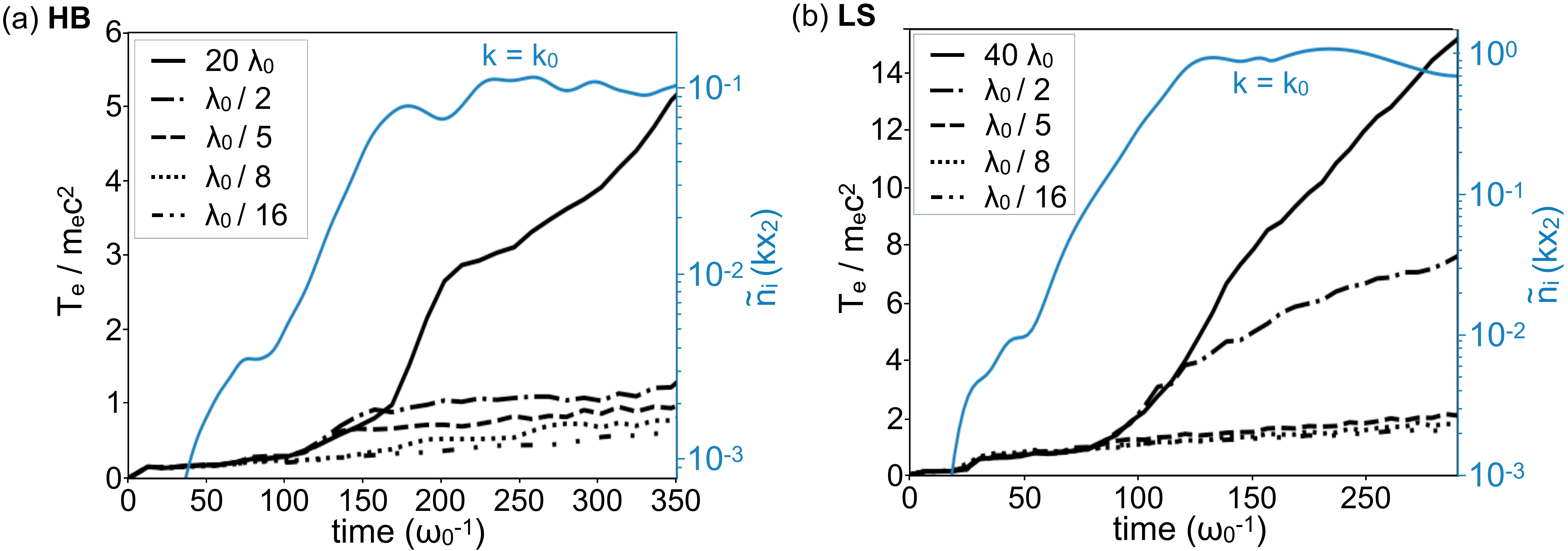}
    \caption{Evolution of electron temperature for different transverse simulation domain sizes (black, left axis). The solid black curves correspond to the simulations in Fig.~\ref{fig:sc} with a transverse box size of $20\,\lambda_0$ for HB (a) and $40\,\lambda_0$ for LS (b) respectively. For these cases the growths of the $k_{x_2} = k_0$ (solid) mode are shown (blue, right axis).}
    \label{fig4.5}
    \end{figure}

To further confirm that the RTI mode with $k_{x_2} = k_0$ is the dominant effect on the onset of strong electron heating and degradation of the ion beam quality, we have performed additional simulations with different transverse domains. We observe that for transverse domain sizes $< \lambda_0/2$, where the dominant RTI mode is prohibited, no significant electron heating is observed. In these cases, the fastest growing high-k modes are still captured and thus can still grow, but $T_e$ remains very low (Fig.~\ref{fig4.5}). For the largest transverse domain size $\gg \lambda_0$, we see that strong electron heating starts near the saturation time of the RTI.

\begin{figure}
    \centering
    \includegraphics[width=.9\textwidth]{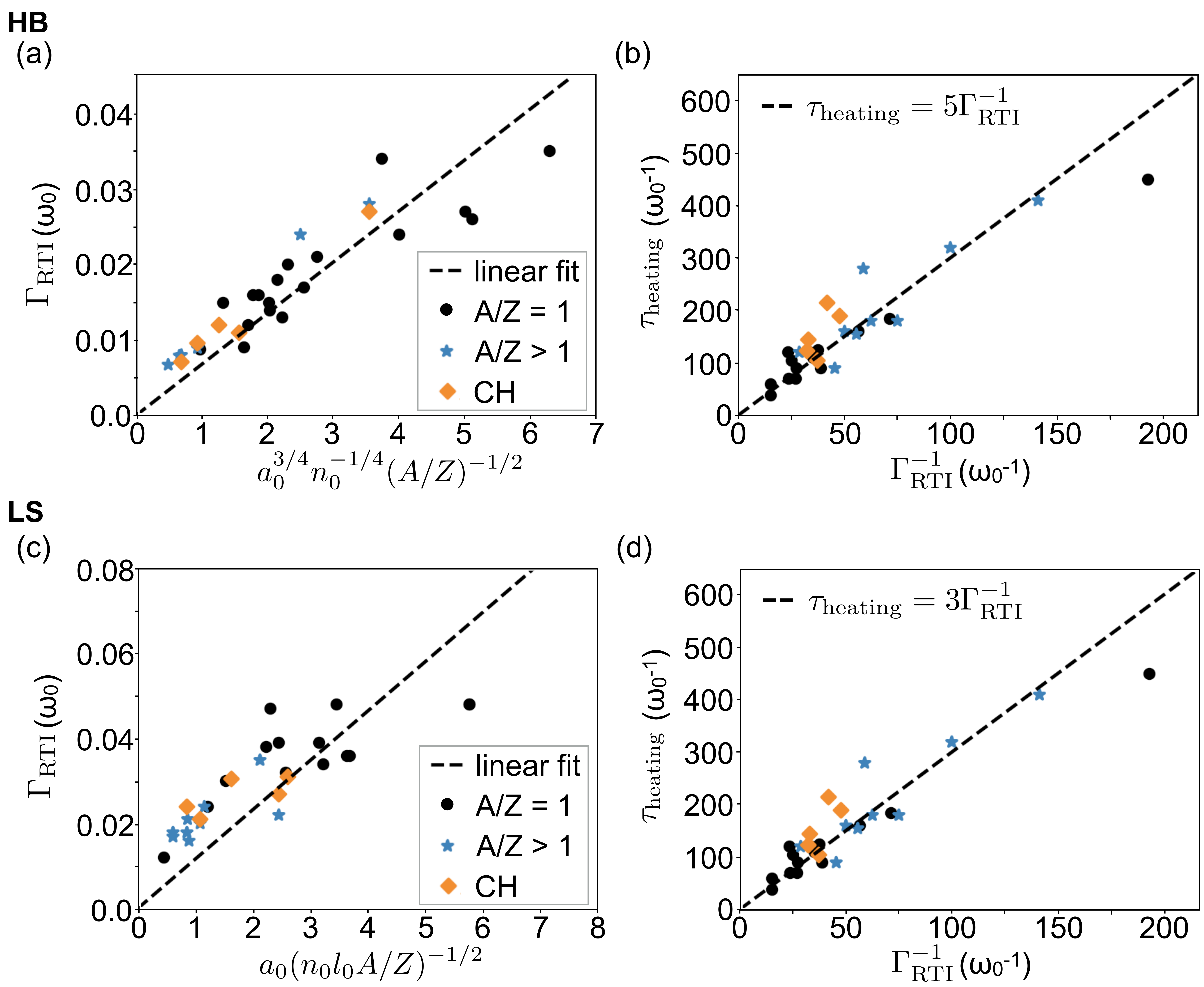}
    \caption{Scaling of the measured growth rate of the $k_{x_2}=k_0$ mode of the surface corrugations [(a),(c)], and strong correlation between the electron heating time $\tau_\text{heating}$ and the growth time of the RTI [(b),(d)], for both HB [(a),(b)] and LS [(c),(d)]. Colored symbols are measurements from 2D PIC simulations.}
    \label{fig:sc_plot}
    \end{figure}

For the laser-driven RTI, the growth rate is $\Gamma_\text{RTI} \propto \sqrt{a_{\rm RPA}k_0}$, where $a_{\rm RPA}=2v_\text{HB}^2/l$ is the acceleration due to radiation pressure (Eq.~\ref{eqn:als} with $R\simeq1$ and $\beta_i\ll1$). For LS, the target thickness is $l_0$, whereas for HB the effective acceleration layer is characterized by the relativistic electron skin depth $l=\sqrt{a_0}c/\omega_{pe}$  \citep{Gamaly1993InstabilityBeam}. Linear fits to the measured growth rates of the $k_0$ mode from the simulations confirm the expected scaling with the following numerical factors (Fig.~\ref{fig:sc_plot}; see also the example fits in Fig.~\ref{fig:sc}c,f): $\Gamma_\text{RTI}[\omega_0]\simeq 0.3a_0^{3/4}\left(n_0[n_c]\right)^{-1/4}\left( Zm_e/(Am_p)\right)^{1/2}$ for HB and $\Gamma_\text{RTI}[\omega_0]\simeq 0.5a_0\left(n_0[n_c]l_0[c/\omega_0] Am_p/(Zm_e)\right)^{-1/2}$ for LS.

We find a strong correlation between $\tau_\text{heating}$ and the growth time of the instability, with $\hat{\tau}_{0,\text{RTI}} \equiv \tau_\text{heating} \simeq 5\Gamma^{-1}_\text{RTI}$ for HB and $\hat{\tau}_{0,\text{RTI}} \equiv  \tau_\text{heating} \simeq 3\Gamma^{-1}_\text{RTI}$ for LS, where $\hat{\tau}_{0,\text{RTI}}$ is defined as the time of the onset of strong electron heating due to the development of the RTI. This is shown in Fig.~\ref{fig:sc_plot}(b) and (d). In order to suppress or significantly mitigate electron heating, the duration of the laser pulse $\tau_0$ should then be smaller than $\hat{\tau}_{0,\text{RTI}}$, which can be written as:

\begin{align}
\hat{\tau}_{0,\text{RTI}} \text{[fs]} &\simeq \begin{cases}
400\, a_0^{-3/4}\lambda_0[\mathrm{\muup m}]\left(\frac{n_0}{n_c}\right)^{1/4}\left(\frac{A}{Z}\right)^{1/2} & \text{ for HB },\\
350\, a_0^{-1}\left(l_0\,[\mathrm{\upmu m}]\lambda_0\,[\mathrm{\upmu m}]\frac{A}{Z}\frac{n_0}{n_c}\right)^{1/2} & \text{ for LS }.
\end{cases} \label{eqn:eng2}
\end{align}

 It is important to note that although several previous works have studied the development of the RTI, the focus had been on developing strategies to mitigate the penetration of the RTI fingers on the accelerated proton species, which included the use of mixed ion species \citep{Yu2010StableFoil,Yu2011SimulationsFoil}, advanced laser configurations \citep{Wu2014SuppressionPulses,Zhou2016ProtonRegime}, and curved targets \citep{Wang2021Laser-IonAccelerator}. These are either challenging to implement in practice (\citealt{Wang2021Laser-IonAccelerator,Zhou2016ProtonRegime}; and have not yet been proven to be effective experimentally), or still lead to significant electron heating \citep{Yu2010StableFoil,Wu2014SuppressionPulses}. As we show here, a quantitative understanding of the detrimental effect that the instability-induced electron heating has on the ion beam quality is critical to produce ion beams with high spectral quality.
 
We should further note that the results presented here have considered only the regime where ions are non-relativistic, which is appropriate for most current and near-future laser systems. In the relativistic regime, the RTI can still grow as shown in previous numerical studies \citep{bulanov2010,sgattoni2014}, but how its growth rate changes and, more generally, its impact on electron heating and the spectral quality of the accelerated ions is not well established. One expects that strong electron heating will still be caused by the development of RTI in the relativistic regime –– Brunel heating will still be present. The resulting space-charge fields will also lead to an increase of the ion energy spread, however the rate at which this happens may be more moderate in the relativistic regime when compared to the non-relativistic case. Furthermore, the resulting expansion of the target will also pose limitations on the acceleration due to the onset of relativistic transparency, as in the cases discussed here. A detailed analysis of the relativistic regime is left for future work.

\subsection{Finite laser spot size}\label{sec:fspot}
The transverse variation of the laser intensity due to its spatial profile naturally leads to non-uniform HB velocities across the surface and results in a change of the surface shape over time, also triggering strong electron heating. The onset time of strong electron heating is approximately when the radiation-pressure-driven displacement $d$ of the target surface is comparable to the laser spot size $w_0$, resulting in a significant change in the local incidence angle. This effect has been previously discussed in the case of normal laser incidence ($\theta_0 = 0^\circ$; \citealt{Klimo2008MonoenergeticPulses,Wan2020EffectsAcceleration}). In fact, this argument can be generalized to other laser incidence angles if we consider the displacement $d$ to be along the axis of the laser resulting 
in a time for electron heating that is independent of $\theta_0$. We define this time as $\hat{\tau}_{0,\text{FS}}$ (``FS'' stands for Finite Spot), and can estimate it as $d(\hat{\tau}_{0,\text{FS}})= w_0$, where for simplicity we assume that the laser intensity is approximately constant in time. 
For HB, we have $d(t)=v_\text{HB}t$. For LS, we consider first that the target experiences a constant acceleration in its rest frame -- \emph{i.e.}, $a_{\rm RPA}\simeq 2v_\text{HB}^2/l_0$ (Eq.~\ref{eqn:als} with $R\simeq1$ and $\beta_i\ll1$). This then gives $d(t)=\left(c\sqrt{c^2+a_{\rm RPA}^2t^2}-c^2\right)/a_{\rm RPA}\simeq \left(cl_0/2v_\text{HB}^2\right)\left(\sqrt{c^2+4v_\text{HB}^4t^2/l_0^2}-c\right)$ (for all cases considered, we have found that the error introduced by taking the non-relativistic ion velocity limit is $\lesssim5$\%). 
Equating $d(\hat{\tau}_{0,\text{FS}})=w_0$ yields

\begin{equation}
\begin{aligned}
\label{eqn:fs}
\hat{\tau}_{0,\text{FS}} \text{[fs]} &\simeq \begin{cases}
200\,a_0^{-1}w_0[\mathrm{\muup m}]\left(\frac{A n_0}{Z n_c}\right)^{1/2}& \text{ for HB }, \\ \noalign{\vskip6pt}
200\,a_0^{-1}\left(\frac{w_0[\mathrm{\muup m}]l_0[\mathrm{\muup m}]A n_0}{Z n_c}\right)^{1/2} & \text{ for LS }.
\end{cases}
\end{aligned}
\end{equation} 

We note that the estimate above considers near diffraction-limited laser focusing where the transverse intensity profile is smooth. If the Strehl ratio is low and speckle-like intensity distributions are present at focus, these will lead to modulations of the surface and laser incidence angle on scales comparable to the speckle size. In that case, in the threshold condition given in Eq. \eqref{eqn:fs} we should replace $w_0$ by the size of the laser intensity speckles, which can be a significant limitation for small-scale ($\sim \lambda_0$) speckles. We further note that with oblique incidence ($\theta_0 > 0$), the ion beam direction will be modified. This can happen due to partial absorption of the laser field, which will impart transverse (along the target surface) momentum to the ions \citep{Macchi2019} and also as the surface is modified to be nearly normal to the laser pulse causing the ion beam direction to be primarily along the laser propagation direction.

\subsection{Dominance of the Rayleigh-Taylor-like instability}\label{sec:4.3}
We now compare the constraints on $\tau_0$ imposed by the RTI via Eq. \eqref{eqn:eng2} and the finite spot size effect via Eq. \eqref{eqn:fs}. RTI dominates when $\hat{\tau}_{0,\text{RTI}} < \hat{\tau}_{0,\text{FS}}$, or equivalently $w_0/\lambda_0\gtrsim 2\left[a_0 n_c/n_0\right]^{1/4}$ for HB, and $w_0/\lambda_0 \gtrsim 3$ for LS. These conditions are met for targets that are relativistically opaque ($a_0 < n_0/n_c$) and for typically used spot sizes $w_0\gtrsim$ (2–3) $\lambda_0$ (assuming near diffraction-limited laser intensity profiles at focus). Thus, we expect that for conditions of interest for laser-driven ion acceleration in overdense targets the surface corrugations by RTI discussed here impose the main limitation on the pulse duration, Eq. \eqref{eqn:eng2}, for high-quality ion beam acceleration.

\section{Laser temporal profile}\label{sec:profile}
    \begin{figure}
    \centering
    \includegraphics[width=.95\textwidth]{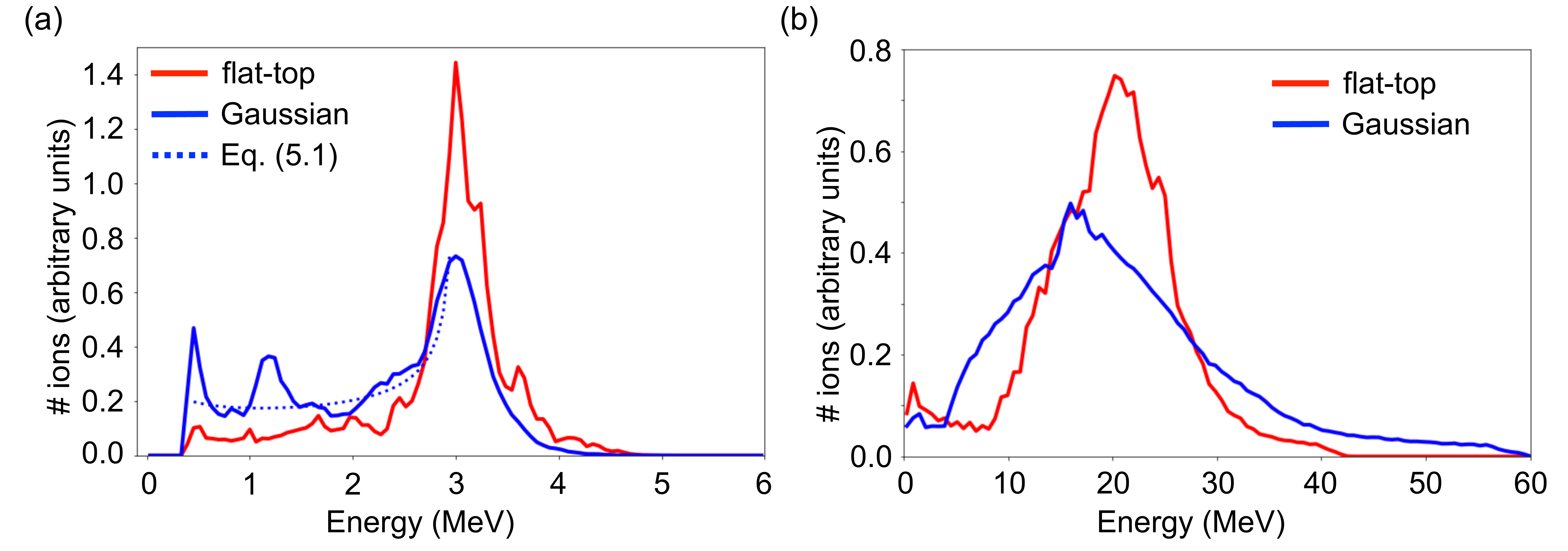}
    \caption{Proton energy spectra from 2D PIC simulations of plane-wave CP laser pulses ($a_0=30$) with either Gaussian (blue) and flat-top (red) temporal profiles normally incident on a target ($n_0 = 136\,n_c$), plotted at the time the laser ends, for (a) HB (with a semi-infinite target) and (b) LS ($l_0 = 0.5\,c/\omega_0$). The Gaussian pulse has $\tau_0= 40\,\omega_0^{-1}$ and the flat-top profile is approximated by a 4th-order supergaussian temporal profile, where the total laser energy is kept the same as the Gaussian pulse. The dotted curve in (a) describes the low-energy component of the spectrum (Eq.~\ref{eqn:temporal}).}
    \label{fig:temp}
    \end{figure}

In this section, we evaluate how different laser temporal profiles impact the quality of the ion beam. This can be particularly relevant for the HB regime, as different ion populations will experience different HB velocities due to $v_\text{HB}(t)\propto a_0(t)$. We have performed simulations in both HB and LS regimes, for a plane wave laser with $\tau_0<\hat{\tau}_{0,\text{RTI}}$, where the laser temporal profile is either Gaussian or a flat-top (approximated by a 4$^\text{th}$-order supergaussian). The total laser energy and $a_0$ is kept the same in both cases. The range of $a_0$, $n_0$, and $l_0$ explored was similar to Sec.~\ref{sec:sc}, and we have varied the pulse duration in the range $0.2\,\hat{\tau}_{0,\text{RTI}}\lesssim \tau_0 < \hat{\tau}_{0,\text{RTI}}$. 

In general, we find that the accelerated proton beams have a relatively narrow energy spread with both profiles, but the energy spread in the Gaussian case is typically larger, by a factor up to $\simeq 2$. Figure \ref{fig:temp} illustrates the typical differences between both profiles for HB and LS. One important feature that we observe in the HB regime is the development of an extended low-energy population ($< $3 MeV in Fig. \ref{fig:temp}a). This is due to the contribution from the low HB velocity phase at the edges of the laser pulse, and can be understood as follows. In a time interval $t$ to $t+dt$, a population of ions $dN\propto n_iv_\text{HB}(t)dt$ is accelerated to a velocity $2v_\text{HB}(t)\propto a_0(t)$ (for $R\simeq1$). Therefore, the resulting energy spectrum will be
\begin{equation}
\begin{aligned}
\label{eqn:temporal}
\frac{dN}{d\epsilon}=\frac{dN}{dt}\frac{dt}{d\epsilon}
\propto\frac{1}{\sqrt{\epsilon\ln(\tilde{\epsilon}_0/\epsilon)}},
\end{aligned}
\end{equation}
for a Gaussian pulse, where $\tilde{\epsilon}_0 = 2 m_i v_\text{HB,0}^2$ is the peak energy of the ions (with $v_\text{HB,0}$ the HB velocity associated with the peak intensity). This matches well the low-energy component of the proton spectrum in Fig.~\ref{fig:temp}(a), which is responsible for the additional energy spread with respect to the flat-top case.

\section{Laser pre-pulse}\label{sec:prepulse}
    \begin{figure}
    \centering
    \includegraphics[width=.95\textwidth]{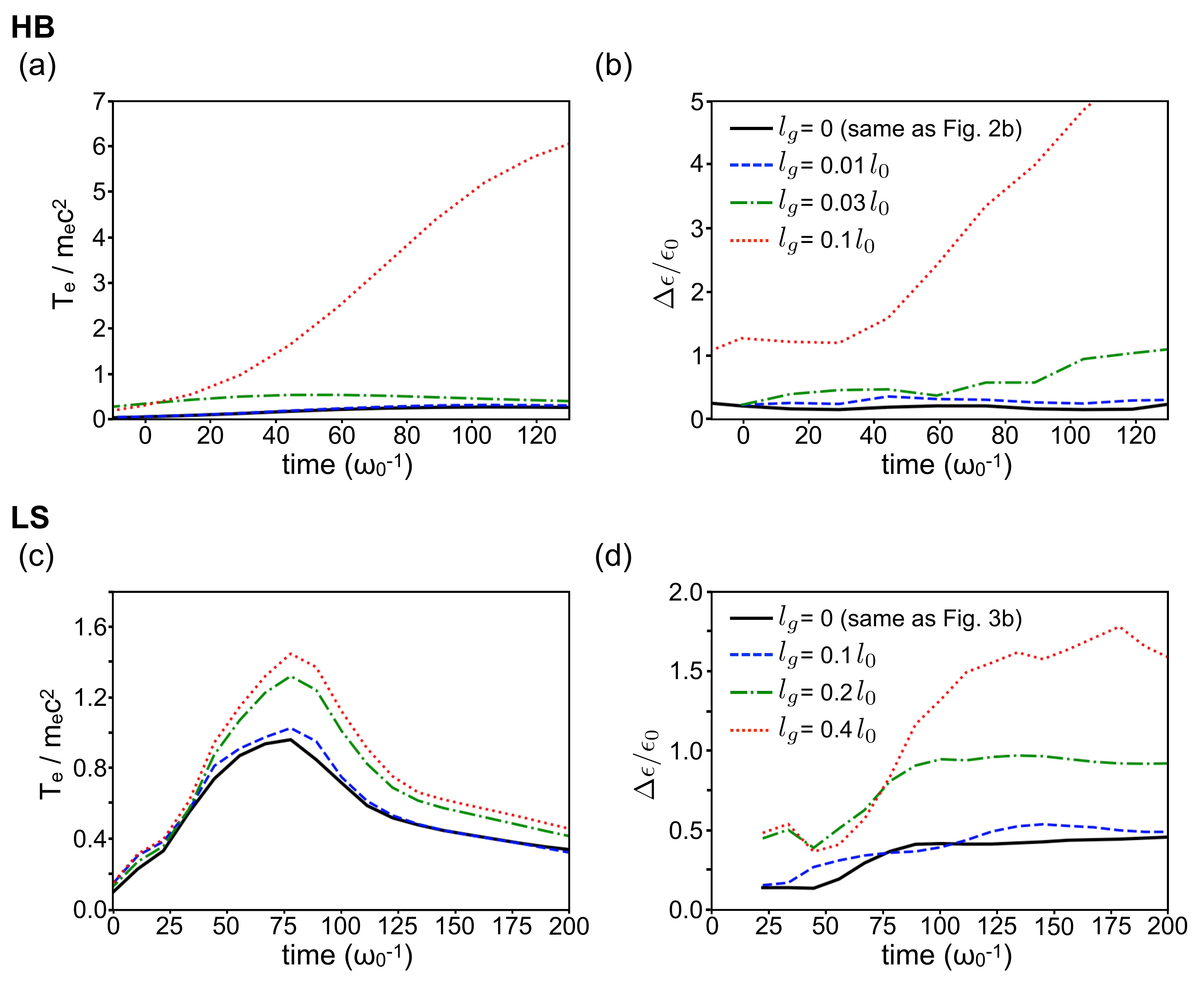}
    \caption{ Results of 2D PIC simulations of the interaction of an intense Gaussian CP laser pulse with a planar target of thickness $l_0$ and an exponential pre-plasma of scale length $l_g$ in the front, in both HB [(a),(b)] and LS [(c),(d)] regimes. [(a),(c)] Temporal evolution of electron temperature $T_e$ and [(b),(d)] ion beam energy spread $\Delta\epsilon/\epsilon_0$. $\Delta\epsilon/\epsilon_0$ is measured for protons within a $10 ^\circ$ opening angle with respect to the laser propagation direction (target normal) and the time $t = 0$ is defined as $\tau_0/2$ before the laser peak intensity reaches the main target.}
    \label{fig:preplasma}
    \end{figure}

In this section, we explore the impact that the development of a pre-plasma induced by a laser pre-pulse can have in triggering early electron heating and affecting the spectral quality of ion beams accelerated via RPA. We consider a pre-plasma with an exponential density profile at the front surface of the target with scale length $l_g$. For the laser and target parameters considered in the short-pulse cases in Fig.~\ref{fig2}(b) and~\ref{fig3}(b), with $\tau_0 = 75\,\omega_0^{-1}$ and $l_0 = 75\,c/\omega_0$ for HB and $\tau_0 = 40\,\omega_0^{-1}$ and $l_0 = 0.085\,c/\omega_0$ for LS, we have performed additional simulations with different levels of $l_g$, which was varied in the range $0.01 \leqslant l_g/l_0 \leqslant 0.4$ (the total target mass is conserved).

Figure~\ref{fig:preplasma} shows that the pre-plasma can significantly impact electron heating and the ion energy spread for levels of $l_g \gtrsim 0.1\,l_0$. In the case of HB, the electron temperature and ion energy spread for a pre-plasma scale length $l_g < 0.1 \,l_0$ ($l_g < 1.2\,\muup$m for $\lambda_0 = 1\,\muup$m) are similar to the case of no pre-plasma, but above this level strong electron heating is triggered, leading to a very fast degradation of ion beam quality even before the pulse arrives at the main target. For LS, for a pre-plasma with $l_g \leq 0.1\,l_0$ the results are also very similar to the no pre-plasma case. However, for $l_g \gtrsim 0.2\,l_0$ ($l_g \gtrsim 2.7$ nm for $\lambda_0 = 1\,\muup$m) both the electron temperature and ion energy spread are observed to increase to nearly twice the values of the no pre-plasma case. These results confirm the need to carefully control the level of pre-plasma to produce ion beams with high spectral quality from RPA and will help inform the laser pre-pulse contrast requirements for future experimental studies.

\section{Optimal regime of radiation pressure acceleration}\label{sec:discussion}

    \begin{figure}
    \centering
    \includegraphics[width=.9\textwidth]{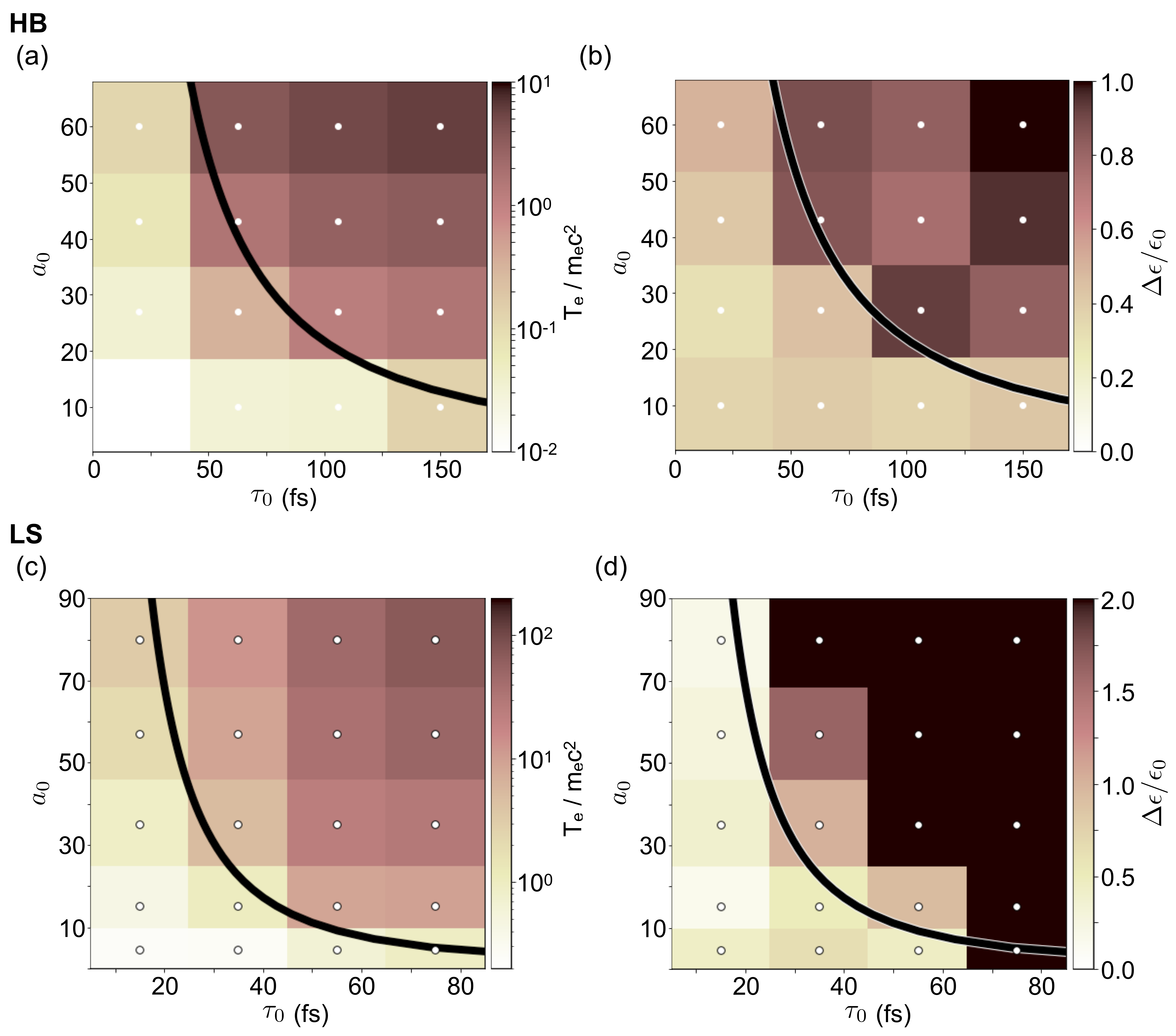}
    \caption{ [(a),(c)] Electron temperature $T_e$ and [(b),(d)] ion beam energy spread $\Delta\epsilon/\epsilon_0$ measured from 2D PIC simulations of a 1$\,\upmu$m wavelength
    Gaussian laser pulse with duration $\tau_0$ and spot size $w_0=7.6$ $\upmu$m interacting with a solid target with $n_0 = 40\,n_c$, $l_0=12\,\muup$m for HB (top row) and $250\,n_c$, $l_0=l_\text{opt}$ for LS (bottom row). $T_e$ is measured at the end of the laser interaction when the maximum is observed. $\Delta\epsilon/\epsilon_0$ is measured for protons within a $10 ^\circ$ opening angle at $t\simeq2\tau_0$. The black curves correspond to the prediction of Eq.~\eqref{eqn:eng2} and the white dots denote the parameters sampled by the simulations.}
    \label{fig:a0t0}
    \end{figure}

Our findings make clear the importance of limiting the pulse duration to control electron heating and obtain quasi mono-energetic ion beams from RPA in both HB and LS regimes. In Fig.~\ref{fig:a0t0} we demonstrate that Eq.~\eqref{eqn:eng2} is robust over a wide range of laser and target parameters even when realistic Gaussian transverse and temporal pulse profiles are considered. We observe that, indeed, the derived threshold condition marks the transition from low to high electron heating (Fig.~\ref{fig:a0t0}a,c) and consequently from low to high energy spread of the accelerated ion beam (Fig.~\ref{fig:a0t0}b,d). By repeating some of the simulations in the high-quality regimes using a small, but finite laser incidence angle, we have also confirmed that in general for an incidence angle $\lesssim10^\circ$, as typically used experimentally, electron heating is still maintained at a low level and the quality of the ion beam remains similar to the case with normal incidence.

For the LS regime, combining the threshold condition for the pulse duration (Eq.~\ref{eqn:eng2}) and the optimal target thickness condition $l_0=l_\text{opt}$ results in a new condition on the maximum laser intensity with important implications for the optimization of ion beam energy spread in LS acceleration \citep{Chou2022}. In particular, we obtain that the laser $a_0$ should be limited by

    \begin{align}
    \label{eqn:a0}
        a_0 \lesssim \hat{a}_0 \equiv \frac{350^2}{\sqrt{2}\pi}\frac{A}{Z}\left(\frac{\lambda_0\text{ [$\muup$m]}}{\tau_0\text{ [fs]}}\right)^2,
    \end{align}

and consequently the maximum energy per nucleon $\epsilon_0$ of the narrow energy spread peak is
    \begin{align}
    \label{eqn:eps0}
        \hat{\epsilon}_0 = m_pc^2\frac{\hat{\xi}^2}{2(\hat{\xi}+1)}, \text{where } \hat{\xi}\simeq 20 \frac{\lambda_0\text{ [$\muup$m]}}{\tau_0\text{ [fs]}}.
    \end{align}
    
In contrast to previous works, and to the common practice of pushing for higher $a_0$ to generate stable LS and higher energy ion beams (\emph{e.g.} \citealt{Qiao2009StablePulses}), Eqs.~\eqref{eqn:a0} and~\eqref{eqn:eps0} indicate the existence of a maximum $\hat{a}_0$ and $\hat{\epsilon}_0$ for high-quality LS ion beams, and show that these are determined primarily by the laser wavelength and pulse duration. In particular, it is worth highlighting that for this optimal regime of acceleration the energy per nucleon of the ion spectral peak does not depend on the target density, composition, and laser energy (transverse spot size). These predictions for the LS regime have been recently validated with 3D simulations in \citet{Chou2022}.

\section{3D simulation results}\label{sec:disc}

    \begin{figure}
    \centering
    \includegraphics[width=.9\textwidth]{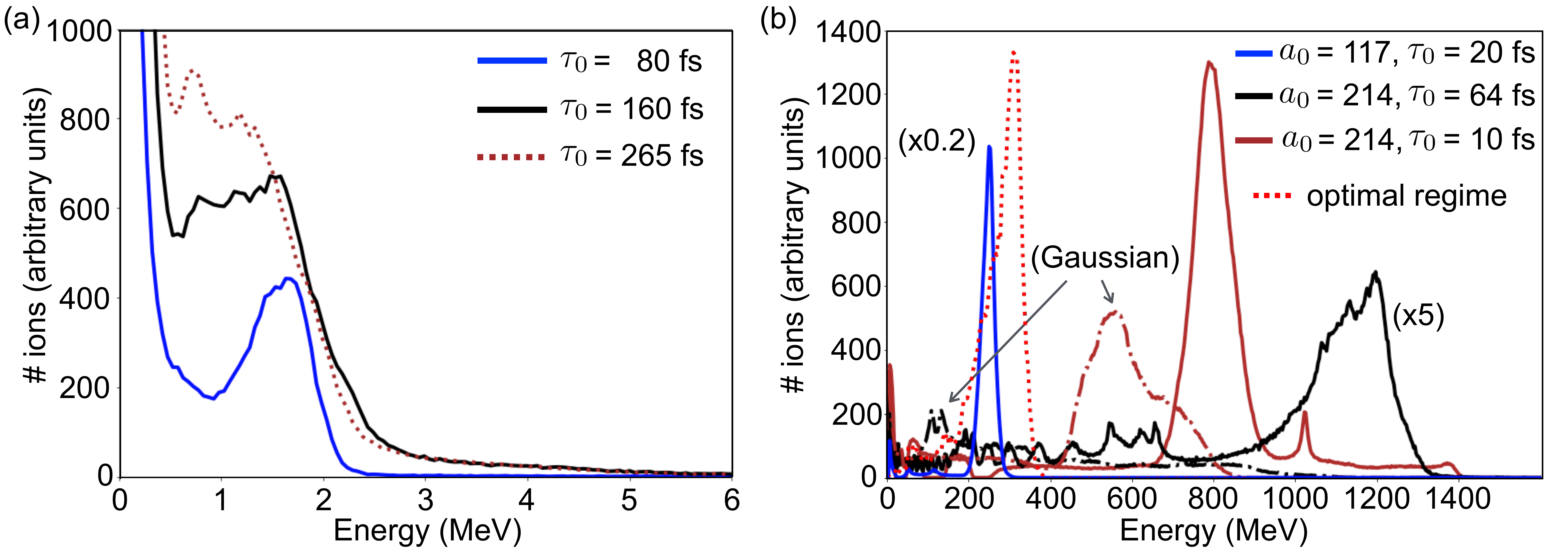}
    \caption{Proton energy spectra from 3D simulations of an intense 1 $\muup$m CP laser irradiating an overdense hydrogen target at normal incidence ($\theta_0=0^\circ$) for different pulse durations. (a) Results for HB regime with laser $a_0 = 12$ and $w_0 = 2\,\muup$m and target $n_0 = 40\,n_c$ and $l_0 = 3.5\,\muup$m, after the proton beam has left the target rear surface ($t \simeq 310$ fs). (b) Results for LS regime with laser $a_0 = 214$ (brown, black) and $a_0 = 117$ (blue), plane wave (solid) or Gaussian transverse profile with $w_0 = 7\,\muup$m (dash-dotted), and target $n_0 = 100\,n_c$ and $l_0 = 350\,$nm, at $t \simeq 2 \,\tau_0$. The dotted red curve shows the spectrum for a $\tau_0 = 15\,$fs laser satisfying Eq.~\eqref{eqn:a0}: $a_0 = \hat{a}_0 = 122$, with $n_0 = 250\,n_c$. All the spectra are measured within a $10^\circ$ opening angle from the laser propagation direction at $t\simeq 2\tau_0$.}
    \label{fig:3d}
    \end{figure}
    
To further explore RPA in more realistic 3D configurations and validate our model for the optimal laser duration for high-quality ion beams, we have performed 3D simulations in both HB and LS regimes. 

In the HB regime, a CP laser with $a_0=12$ ($I \simeq 2\times10^{20}$ W/cm$^2$ for $\lambda_0 = 1 \muup$m), $w_0 = 2\,\muup$m is incident with $\theta_0 = 0^\circ$ on a planar electron-proton target with $n_0 = 40\,n_c$ and thickness of $3.5\,\muup$m, corresponding to the conditions of typical liquid hydrogen jet targets \citep{Kim2016DevelopmentExperiments,Gauthier2017HighJets,Gode2017RelativisticInteractions,Obst2017EfficientJets}. A 4$^\text{th}$-order supergaussian temporal profile is used. For these parameters, Eq. \eqref{eqn:eng2} indicates $\tau_0 \lesssim 160\,$fs for high-quality proton beams to be produced. We thus run three simulations with pulse durations of $\tau_0=80$, 160, and $265\,$fs to test this criterion. In Fig.~\ref{fig:3d}(a), we show the energy spectra of the proton beams exiting the target from the rear surface for these three cases. When the pulse duration is smaller than the predicted threshold, a quasi mono-energetic proton beam is generated, peaking at roughly $1.8\,$MeV, which corresponds to $v_i=2v_\text{HB}\simeq 0.031\,c$ and is consistent with Eq.~\eqref{eqn:vhb} for $R\simeq1$. When the pulse duration is comparable to the threshold, we observe that the ion spectral peak at similar energy is still visible, but less prominent. For the longer pulse duration ($\tau_0=265\,$fs) we find substantial electron heating ($T_e\simeq 3\,$MeV), and a strong TNSA field is generated at the rear surface, which broadens the proton energy spread. No clear spectral peak is observed at the same energy. These results confirm the validity of the derived limit on pulse duration for HB based on the development of surface corrugations and associated electron heating.

We note that the generation of narrow energy spread HB ion beams with similar peak energies around 1 MeV have been reported in \citet{Palmer2011MonoenergeticShock}, where the laser pulse duration appears to meet the criterion of our model. Specifically, they considered a CP CO$_2$ laser with $\lambda_0 = 10\,\muup$m, $a_0\simeq 0.7$, $w_0 = 70\,\muup$m, and a hydrogen gas jet target with $n_0\lesssim 10\,n_c$, $l_0\sim800\,\muup$m. The pulse duration used $\tau_0 \simeq 6\,$ps was less than $\hat{\tau}_{0,\text{RTI}}\simeq 9\,$ps as required by Eq.~\eqref{eqn:eng2}. The measured ion beam energies were also shown to be consistent with efficient HB.

For the LS regime, additional 3D simulations have also been performed to illustrate the change in the ion beam quality for different pulse durations. We simulate a CP laser impinging on a planar electron-proton target with $n_0 = 100\,n_c$ and $l_0 = 350\,$nm at normal incidence. The laser pulse is temporally Gaussian and the transverse profile is either Gaussian with $w_0 \simeq 7\,\muup$m or plane wave. These parameters are similar to those used in \citet{Qiao2009StablePulses}, where the generation of a GeV proton beam was obtained in 2D simulations and it was argued that extreme laser intensities were needed to obtain stable high-quality ion beams. Two cases were illustrated: unstable acceleration with $a_0 = 117$ and a stable regime with $a_0 = 214$. Under these conditions, Eq. \eqref{eqn:eng2} predicts that high-quality proton beams require $\tau_0 \lesssim20\,$fs for $a_0 = 117$ and $\tau_0\lesssim10\,$fs for $a_0 = 214$, both of which are much smaller than the duration $\tau_0\simeq 64\,$fs used in the cited work.

Fig.~\ref{fig:3d}(b) shows the comparison of the proton spectra at the same time as Fig. 4 in \citet{Qiao2009StablePulses}, which corresponds to $t\simeq 2\tau_0$ after the laser interaction. For the case with $a_0=117$, when a pulse duration $\tau_0 = 20\,$fs is used, we observe the generation of a high-quality proton beam with $250\,$MeV, $\Delta \epsilon/\epsilon_0 \simeq 13$\%. For the same parameters but using $\tau_0$ = 64 fs (not shown) we observe the generation of a beam with $460\,$MeV, $\Delta \epsilon/\epsilon_0 \simeq 32$\%, similar to their green curve. This indicates that it is the onset of electron heating associated with the surface instability in the longer pulse that leads to the increase of the energy spread and to a significant reduction of the coupling efficiency.

For the highest intensity case ($a_0=214$), we observe that similarly the proton beam energy spread is improved from $\Delta \epsilon/\epsilon_0 =$ 20\% with $\tau_0=64\,$fs to $\Delta \epsilon/\epsilon_0 =$ 12\% with $\tau_0=10\,$fs. We further observe that for a transverse Gaussian laser profile the importance of the short pulse duration is even more dramatic. Defining the laser-to-proton energy conversion efficiency $\eta$ as the fraction of the laser energy being carried by the beam within a $10^\circ$ opening angle, for $\tau_0 = 10\,$fs we obtain high-quality proton beams with $\sim 600\,$MeV, $\Delta\epsilon/\epsilon_0 = $30\%, and $\eta =$ 27\%, whereas with $\tau_0 = 64\,$fs, we have $\sim 100\,$MeV, $\Delta\epsilon/\epsilon_0 = $45\%, and $\eta =$ 0.5\%. The proton beam energy obtained with the short pulse ($\tau_0 = 10\,$fs) is comparable to the prediction from 1D theory of 700 MeV (Eq.~\ref{eqn:beta}). We note that this is slightly higher than the prediction of Eq.~\eqref{eqn:eps0} because the target thickness considered in \citet{Qiao2009StablePulses} was $l_0\lesssim l_\text{opt}$. More importantly, these results confirm that the process that controls the stability and spectral quality of the LS accelerated ion beams is the electron heating via the development of RTI at the target surface and that choosing the appropriate pulse duration and laser intensity is critical for the acceleration of high-quality ion beams. 

Figure~\ref{fig:3d}(b) also includes the results of a 3D PIC simulation for which the laser and target parameters satisfy the optimal LS regime (Eq.~\ref{eqn:a0}). We have chosen a set of parameters relevant for near-future short-pulse laser facilities, where a Gaussian laser pulse profile with $\tau_0 = 15\,$fs, $w_0 = 5\,\muup$m, and $a_0 = \hat{a}_0 \simeq 122$ according to Eq.~\eqref{eqn:a0} is used. The target has $n_0 = 250\,n_c$ and $l_0 = l_\text{opt}$. Under this optimal regime, we indeed observe stable acceleration of the protons via LS leading to the generation of a narrow energy spread proton beam with peak energy $\epsilon_0 \simeq 310\,$MeV in very good agreement with the prediction of Eq.~\eqref{eqn:eps0}, $\Delta \epsilon/\epsilon_0 \simeq$ 25\%, and $\eta\simeq 2\%$. The total laser-to-proton energy conversion efficiency into $4\pi$ is $\sim20\%$.

\section{Conclusions}\label{sec:con}
In summary, we have presented a detailed study of electron heating and ion acceleration in the interaction of intense laser pulses with overdense plasmas, with a particular focus on radiation pressure acceleration. We have shown that electron heating controls the interplay between different ion acceleration mechanisms. For circularly polarized lasers, the onset of strong electron heating is dominated by the development of corrugations of the target surface on the scale of the laser wavelength due to the laser-driven RTI. In the HB regime, electron heating leads to the development of a collisionless shock and the transition to CSA and TNSA. In the LS regime, electron heating causes fast target expansion and onset of relativistic transparency leading to a significant increase of the ion energy spread. We have shown that to reduce or suppress electron heating and obtain high-quality (low energy spread) ion beams by RPA it is critical to use laser pulses much shorter than the saturation time of the RTI. Using 2D and 3D PIC simulations, we have demonstrated that when such short pulses are used high-quality proton beams can be produced with maximum energy comparable to the optimal 1D RPA theory, even for lasers with Gaussian transverse and temporal intensity profiles.

Interestingly, we note that a few of the previous experimental studies that reported low energy spread ion beams from either the HB \citep{Palmer2011MonoenergeticShock} or LS \citep{Henig2009Radiation-PressurePulses,Steinke2013} acceleration appear to meet the laser duration criteria developed in this work. This is encouraging and the understanding provided by our work can help guide future experimental developments in this area. For example, for the parameters of high-power, high-contrast state-of-the-art and near future laser systems with $\tau_0 \simeq 15\,$fs, such as the ELI-NP \citep{Doria2020}, Apollon 10 PW \citep{Papadopoulos2016}, and EP-OPAL \citep{EP-OPAL} facilities, our 3D PIC simulations demonstrate the possibility to produce $\sim 300\,$MeV proton beams with $\sim 25\%$ energy spread and high coupling efficiency.

\hfill \break
\textbf{Acknowledgments}

The authors thank the OSIRIS Consortium, consisting of UCLA and IST (Portugal) for the use of the OSIRIS 4.0 framework and the visXD framework. Simulations were performed at Cori (NERSC) and Theta (ALCF) through ERCAP and ALCC computational grants.

\hfill \break
\textbf{Funding}

This work was supported by the U.S. Department of Energy SLAC Contract No. DEAC02-76SF00515, by the U.S. DOE Early Career Research Program under FWP 100331, and by the DOE FES under FWP 100182.

\hfill \break
\textbf{Declaration of interests}

The authors report no conflict of interest.

\bibliographystyle{jpp}
\bibliography{ref}

\end{document}